\documentclass[10pt]{iopart} 
\usepackage{a4wide}
\usepackage{epsfig,multicol}

\def\be{\begin{equation}}
\def\ee{\end{equation}}
\def\bea{\begin{eqnarray}}
\def\eea{\end{eqnarray}}

\usepackage{iopams}
\usepackage{setstack}

\begin{document}
\hfill AEI-2007-155
\\ \\ \\
\title{Is there a tower of charges to be discovered?}

\author{T. M\aa nsson}

\address{Max-Planck Institut f\"ur Gravitationsphysik,
Albert-Einstein-Institut \\
Am M\"uhlenberg 1, D-14476 Potsdam, \\ Germany}

\ead{teresia@aei.mpg.de}
 
\begin{abstract}
We investigate higher loop integrability for a $q$-deformation of the
 $\mathfrak{su}(2)$-sector of $\mathcal{N}=4$ SYM theory. First we
construct a generalisation of the long range spin chain, which for the
lowest orders describes the non-deformed dilatation operator. This
generalised model is built up from Temperley-Lieb algebra generators
and describes the deformed theory to at least two loops. When constructing
the model we have demanded the existence of one commuting charge, which
 puts strong constraints on the parameters to three loop orders. We also
write down the five first
charges for this model at two loops order.

 Our main goal is to obtain an
explicit expression for an infinite number of commuting charges, all
commuting with the dilatation operator. This would imply integrability. As
a step towards this goal we present in this paper an expression for
a generic local charge of the one-loop dilatation operator, which
happens to be a generator of the Temperley-Lieb algebra. 
\end{abstract}



\section{Introduction}
One decade ago, Maldacena conjectured the famous AdS/CFT correspondence
relating a strongly coupled conformal field theory (in particularly
the so-called  $\mathcal{N}=4$ supersymmetric Yang-Mills theory)
with a weakly coupled string theory on an Anti-de Sitter
background and vice versa. The energies of the strings are mapped to the
anomalous dimensions of certain single trace operators 
(composite gauge invariant operators) in the
field theory. If the duality was correct it would  enable us to calculate
quantities in either theory which were previously almost unattainable.

Integrability has shown up in a remarkable way in the context of
 $\mathcal{N}=4$ supersymmetric Yang-Mills theory with the gauge group
$SU(N_c)$, when calculating the anomalous dimensions of single trace
operators in the large $N_c$-limit (the planar limit). In particular, it
has been pointed out that the dilatation operator, which is defined to give
the anomalous dimensions as its eigenvalues, can,
 to one loop order, be mapped
to an integrable spin chain \cite{Minahan:2002ve,Beisert:2003yb}.
Furthermore, in some sectors in the planar limit of the theory, the
dilatation operator has been proven to be integrable to a few loops order.
In particular, for the $\mathfrak{su}(2)$ sector the dilatation operator is
now known to four orders accuracy
\cite{Beisert:2003tq,Bern:2006ew,Beisert:2007hz}. The exact
expression for the operator is not yet clear. For the $\mathfrak{su}(2)$
sector it has been
shown  \cite{Serban:2004jf} that the first few orders agree with the
Inozemtsev spin chain \cite{Inozemtsev:2002vb}, and it
can also be obtained as a limiting case of the BDS-Hubbard model
\cite{Rej:2005qt}. It is believed that the theory is integrable to all
loops order in all sectors. A conjectured all loop asymptotic Bethe Ansatz
has been written down \cite{Beisert:2004hm,Arutyunov:2004vx,Beisert:2005fw}.

Deformations of the theory are of physical interest. Integrability has been
studied for some deformations preserving the conformal structure of the
theory, in particular the $q$-deformation (or Leigh-Strassler deformation)
\cite{Roiban:2003dw,Berenstein:2004ys,Beisert:2005if,Freyhult:2005ws,
Bundzik:2005zg,Mansson:2007sh}.
There it has been found that the one-loop dilatation operator in the planar
limit is integrable for specific values of the parameters and in
particular sectors, but beyond one loop not much is known
\cite{Frolov:2005ty}.

It is an interesting question if there exists a natural generalisation
of the integrable all-loop model for the $\mathfrak{su}(2)$ case,
 to the $q$-deformed
 $\mathfrak{su}(2)$-sector. By $q$-deformed, we now mean the $q$-deformed
Leigh-Strassler deformation of $\mathcal{N}=4$ supersymmetric Yang-Mills
theory \cite{Leigh:1995ep}.
 In practise, that means that the operator $I-P$ should be
exchanged with the Temperley-Lieb generators $e_i$ in the Feynman diagrams
which give rise to the dilatation operator. These types of generalisations
appear to have been fruitful before, e.g. the model
discussed in \cite{Jacobsen}.

The spin chain Hamiltonian, to which the dilatation operator is mapped,
can in the non-deformed $\mathcal{N}=4$ SYM case be written as a loop
expansion\cite{Beisert:2003tq,Serban:2004jf},
$$
\label{Hamiltonian:nondeformed}
 H=L+\sum_{k=1}^\infty\left(\frac{\lambda}{16\pi^2}\right)^k h_k\,,
$$
where the first three $h_k$ are
$$
h_2=-6L+8\sum_i P_{i,i+1}-2\sum_i\{P_{i,i+1},P_{i+1,i+2}\}
$$
\bea
\fl h_3=60L+\sum_i -104 P_{i,i+1}+24\{P_{i,i+1},P_{i+1,i+2}\}
+4 P_{i,i+1}P_{i+2,i+3} \nonumber \\
-4(P_{i,i+1}P_{i+1,i+2}P_{i+2,i+3}-P_{i+2,i+3}P_{i+1,i+2}P_{i,i+1})\,,
\eea
where $L$ is the length of the trace operator in the
field theory, or alternatively the length of the spin chain.

The lowest known orders of the dilatation operator for the non-deformed
theory satisfy perturbative integrability ($K$ is the number of known
orders):
$$
Q_l=\sum_{k=0}^{K} \lambda^k Q_l^{(k)},
$$
$$
[Q_l,Q_m]=\mathcal{O}(\lambda^{(K+1)}).
$$
Up to the first few orders, the Hamiltonian agrees with an effective
spin chain model obtained as a certain limit of the Hubbard model. We would
like to see if this is also the case for our generalisation of the model.

We have two main motivations for this paper, one is to clarify whether
the $q$-deformed $\mathfrak{su}(2)$ model can be integrable to higher
loops, and
secondly on a very general basis investigate the possibility to 
generalise the type of long-range integrable spin chain 
describing
the dilatation operator in the $\mathfrak{su}(2)$ sector in $N=4$ SYM
theory,
to integrable long range systems built up with Temperley-Lieb generators.
We will start by writing down the simplest long-range model to four orders
in $\lambda$ which to three loops order reduces in the $q\rightarrow 1$ 
limit to the one above, modulo commuting charges. We would like to show
that this generalised model is indeed integrable. Our plan is to prove the
existence of an infinite number of commuting charges, by writing down
an explicit expression for them. In order to do so we would first need to
have a generic expression for any local charge commuting with the first
order charge. The first order charge
can be represented by a generator $e_l$ of the Temperley-Lieb algebra.

In \cite{Grabowski:1994ae,Anshelevich:1980da},
 the authors derived expressions for a
generic charge of the Heisenberg spin chain. In \cite{Grabowski:1994qn},
it was shown that the Hubbard model doesn't have a boost operator. This
makes it harder to derive the general charge. One motivation
for them to 
 understand the structure of the charges
for the Heisenberg spin chain, was that it 
could reveal the structure of the charges of
the Hubbard model, and also of long range spin chains which has the
Heisenberg spin chain as a limiting case, i.e.
the Haldane-Shastry spin chain.

We think  that the underlying structure of the charges
at one loop will be related in the same way to a deformed Hubbard model or
Haldane-Shastry model. 
According to our knowledge, very little is known about long
range integrable models and we expect this generalisation to shed some
light into this area.

\section{The $q$-deformed $\mathfrak{su}(2)$ sector}
In the case of the $q$-deformed $\mathfrak{su}(2)$ sector, to one loop
accuracy, the action of the dilatation operator on two adjacent scalar
fields in the trace operator can be described by either the action of the
XXZ spin-1/2 chain (when considering the
sector with one holomorphic and one anti-holomorphic scalar field
\cite{Mansson:2007sh}), or a 2d representation of the Temperley-Lieb algebra 
(when considering the gauge theory sector with two holomorphic, or two
anti-holomorphic scalar fields \cite{Roiban:2003dw,Mansson:2007sh}).
The action of $e_i$ on two neighbouring spin sites $i$ and $i+1$, in the
latter case, can be can be represented by the matrix
\be
e_i=
\left(\begin{array}{cccc}
0 &0 & 0 & 0 \\
0 &|q| & -e^{i\beta} & 0 \\
0 & -e^{-i\beta} & |q| & 0\\
0 & 0 & 0 & 0
\end{array}\right),
\ee
where $q=|q|e^{i\beta}$ ($\beta\in\mathbb{R}$) is the deformation
parameter, which deforms the superpotential in the field theory.

In the spin chain description, we consider closed chains since these
correspond to trace operators in the field theory. The difference between
the XXZ spin chain Hamiltonian and the Temperley-Lieb generator vanishes
in the case of closed spin chains. Thus we write the one-loop dilatation
operator as a sum of generators of the Temperley-Lieb algebra,
\be
\label{temperley-hamiltonian}
Q_1=\sum_{l=1}^L e_l\,,
\ee
where $L$ is the length of the spin chain, which is periodic. The
generators $e_i$ act on two adjacent spin sites $i$ and $i+1$, and
fulfill the algebraic relations
\bea
&e_ie_{i\pm 1}e_i=e_i &\qquad \nonumber \\
&e_i e_j=e_je_i       &\qquad,|i-j|\geq 2 \label{Temperley}\\
&e_i^2=\gamma e_i     &\qquad,\gamma=|q|+|q|^{-1}\,. \nonumber
\eea

We should point out that even though we are looking at this system with
particular interest in the representation of the deformed
$\mathfrak{su}(2)$-sector, everything that follows is much more general,
and is independent of the choice of representation. Only the fundamental
Temperley-Lieb algebra above is used in the sequel.

Now we would like to start discussing what is our real interest, the higher
loop contributions to the dilatation operator.
\emph{Up to one loop it is clear that due to the periodicity
it does not matter which two scalar fields we choose as our representation.
At higher loop order we get contributions consisting of products of these
main building blocks, and in principle we should check whether
perodicity implies that the difference term vanishes also for these
products.}

For every increasing loop order, we have one more site involved in the
interaction, as
can be seen from the structure of the Feynman diagrams and the fact that we
are taking the planar limit. At one loop order, there are only nearest
neighbour interactions, at two loops there are three adjacent spin
sites involved, and so on. In addition, the interaction can be expressed
in terms of products of the generators $e_i$, and they always come in
pairs with another product with the ordering reversed, i.e. the interaction
operator $e_{1}e_4e_{3}e_{2}$ will be accompanied by the operator
$e_{2}e_{3}e_4e_{1}$, with the same coefficient in front.

The most general expression satisfying these conditions to three loops,
modulo commuting charges and linear combinations of the charge itself
multiplied by constants times $\lambda$, is
\bea
Q_1&=\lambda\big[e_i+\lambda \{e_i,e_{i+1}\}+
\lambda^2\left(\alpha_1\{e_i\{e_{i+1},e_{i+2}\}\}+
\alpha_2 \{e_i,e_{i+2}\} \right) \nonumber \\
&\quad\lambda^3\left(\{e_i\{e_{i+1},\{e_{i+2},e_{i+3}\}\}\}+
\ldots \right)\big]+\mathcal{O} (\lambda^5)
\eea
Our four-loop term is not the most general possibility. At four
loops order we will at the moment not take interest in the most general
expression. We will, however, use a form with maximum interaction range,
and include the total anti-symmetrization of the four adjacent spin sites.

Our basic question now is whether this type of model is integrable, and for
which parameter values. All we know from the outset is that it is
integrable when  $q\rightarrow 1$. Thus we know which parameter
values are needed in that limit. A condition for integrability is that
there must exist a $Q_2^{(2)}$ such that
\be
\label{demand}
[Q_1^{(1)},Q_2^{(2)}]=[Q_2^{(2)},Q_1^{(1)}]\,,
\ee
assuming that the charge number $k$ is perturbatively expressed as
\be
Q_k=\sum_{\lambda=1}^\infty Q_k^{(l)} \lambda^l\,.
\ee
The condition (\ref{demand}) comes from demanding perturbative
integrability. Before going on to write down the charges that fulfill this
condition, let us introduce some useful notation.

All expressions we will deal with will be combinations of commutators and
anti-commutators of the Temperley-Lieb generators $e_i$. Therefore, it will
be convenient for us to introduce the following notation,
\be
e_{i_1,i_2,\bar{i}_3,i_4,\bar{i}_5,i_6}=
[e_{i_1}[e_{i_2}\{e_{i_3}[e_{i+4}\{e_{i_5},e_{i_6}\}]\}]]\,,
\ee
which naturally reduces to a Temperley-Lieb generator when there is only
one index. We also like to define a special ``index'' $B_i$, which we
define to mean
$$
e_{B_i}:=e_ie_{i-1}e_{i+1}e_i\,.
$$
We would like to introduce some more notation for future convenience.
We will encounter terms of the form
\be
e_{\widetilde{i}_1,\widetilde{i}_2,\ldots,\widetilde{i}_k,\ldots,\widetilde{i}_l},
\qquad  \mbox{where}\qquad i_k<i_{k+1}\,,
\ee
where the tilde means that the index can be either barred or not. It is
practical to have a special word for the number $i_l-i_1+1$, where $i_1$ is
the lowest index $i_1$ of the term and $i_l$  the highest index. Let us
call it ``range number''. We also need a special word for the total number
of gaps. If e.g. $i_k=i_{k+1}+2$, that would be a gap of one. The total
number of gaps is the sum of all $g_k$, where $i_k=i_{k+1}+1+g_k$. We can
characterise the terms by their range number, number of bars
(anti-commutators), and total number of gaps.

We now introduce the notation
\be
T^N_{k,j}\,,
\ee
which will be a sum of terms with range number $N$, number of bars $k$, and
total gap number $j$. As an example,
\bea
T^5_{2,1}&= \sum e_{\overline{i+1},\overline{i+3},i+4,i+5}
+e_{\overline{i+1},i+3,\overline{i+4},i+5}
+e_{\overline{i+1},\overline{i+2},i+4,i+5} \nonumber \\
&\quad{}+e_{i+1,\overline{i+2},\overline{i+4},i+5}
+e_{\overline{i+1},i+2,\overline{i+3},i+5}
+e_{i+1,\overline{i+2},\overline{i+3},i+5}\,.
\eea
We also introduce the notation $TB^N_{l,k}$, which is the same as
$T^N_{l,k}$ but with one of the indices being the symbol $B_i$, e.g.
\bea
TB^5_{1,0}&=\sum e_{\bar{B}_{i+1},i+3,i+4}+
e_{B_{i+1},\overline{i+3},i+4}
 +e_{\bar{i},B_{i+2},i+4}+
e_{i, \bar{B}_{i+1},i+3}+ \nonumber \\
& e_{\bar{i},i+1,B_{i+3}}+
e_{i,\overline{i+1},B_{i+3}}\,.
\eea

Now we are ready to use our notation to write down two charges fulfilling
restriction (\ref{demand}). After demanding the existence of one
commuting charge, the coefficients $\alpha_1$ and $\alpha_2$ are
completely determined. To fourth order we have several options, and here we
have just chosen one possibility:
\bea
Q_1&=T^1_{0,0}+\lambda  T^2_{1,0}+
\lambda^2\left(-\frac{2}{ \gamma} T^3_{1,1}+ T^3_{2,0}\right) 
+\lambda^3\left(
\left(-1+\frac{2}{\gamma}-\frac{2}{\gamma^2}\right)T^3_{1,1}\right.
 \nonumber \\
&\quad{}\left.-\frac{2}{\gamma}T^3_{2,0}+
\left(\frac{2}{\gamma}-\frac{10}{\gamma^2}\right)T^4_{1,2}
-\frac{2}{\gamma} T^4_{2,1}+2TB^{3}_{0,0}+  T^4_{3,0}\right),
\label{Hamiltonian2}
\eea
and the second charge,
\bea
Q_2&=T^2_{00}+\lambda T^3_{1,0}+
\lambda^2\left(-\frac{2}{\gamma}(T^4_{1,1}+T^3_{1,0})+T^4_{2,0}\right) 
+\lambda^3\left(-(1-\frac{4}{\gamma}+\frac{12}{\gamma^2})T^4_{1,1}
 \right.\nonumber\\
&\quad{}
\left.-(4-\frac{2}{\gamma}+\frac{2}{\gamma^2})T^3_{1,0}
 +(\frac{2}{\gamma}-\frac{10}{\gamma^2})T^5_{1,2}
-\frac{2}{\gamma}(T^5_{2,1}+2 T^4_{2,0})+2TB^4_{0,0}+
T^5_{3,0} \right). \nonumber
\eea

From this little exercise we can conclude that to second loop order we are
totally free, and the parameter values we would get from a field
theory calculation don't matter. But to third order they are
essential. Thus, by doing a three-loop field theory calculation
 we could exclude integrability if the coefficients do
not come out as above. We also see that we have a $q$-dependence in the
coefficients. It would be quite amazing if the field-theoretical
calculation turned out to give exactly this dependence.

In cases where a boost operator that generates all the charges exists, it
is enough to show that the first and second charges commute. It then
follows that all the rest will commute. For the one-loop Hamiltonian, the
boost operator is well known \cite{Tetelman:1982}
\be
\label{boostoperator}
B=\sum_{l=-\infty}^{\infty} l e_l\,.
\ee
Note that the boost operator is only well-defined for the infinite spin chain,
but this is not a problem. Unfortunately, when trying to find any boost
operator for the higher-loop dilatation operator, we run into the same
problem as in the case of the Hubbard model with a factor of two 
appearing, which destroys the Ansatz for the boost operator
from working \cite{Grabowski:1994qn}.

In order to extract the generic charge commuting with our Hamiltonian, we
first need an explicit expression for the charges in the one-loop
case. In the next section we will derive a simple expression for a 
generic local charge of the Hamiltonian (\ref{temperley-hamiltonian}).

\section{The infinite tower}
The fact that the charges can be generated using the boost operator
(\ref{boostoperator}) gives us information about their algebraic structure.
We claim that they can be expressed purely in terms of the 
quantities $T^N_{k,l}$ defined in the previous section.

We will now provide an induction proof for this claim. As seen in
the previous section, this is already true for the lowest charges. We now
prove it for the charge $Q_{N+1}$, assuming that it is true for $Q_N$.
Due to the existence of the boost operator (\ref{boostoperator}), the set
of charges can be generated recursively as follows,
\be
Q_{N+1}=[B,Q_N]\,.
\ee
Now we consider a generic term in the $Q_N$ charge, and study the result of
commutation with the boost operator. The following relations are almost all
we need to know in order to calculate commutators. First the relations
involving one anti-commutator, i.e. one barred index:
\bea
&&e_{2,1,\bar{2},3}=
e_{\bar{2},3}-e_{\bar{1},2}-\frac{\gamma}{2} e_{\bar{1},\bar{2},3}
+\frac{\gamma}{2} e_{1,2,3}+2\,e_{B_2} \label{eq:oneanti1} \\
&&e_{2,\bar{1},2,3}=
e_{\bar{2},3}-e_{\bar{1},2}+\frac{\gamma}{2} e_{\bar{1},\bar{2},3}
-\frac{\gamma}{2} e_{1,2,3}-2\,e_{B_2} \label{eq:oneanti2} \\
&&e_{\bar{2},1,2,3}=
e_{\bar{2},3}+e_{\bar{1},2}-\frac{\gamma}{2} e_{\bar{1},\bar{2},3}
+\frac{\gamma}{2} e_{1,2,3}-2\,e_{B_2}\,.\label{eq:oneanti3}
\eea
Then the relations involving even number of anti-commutators:
\bea
&&e_{2,1,2,3}=e_{2,3}-e_{1,2}-\frac{\gamma}{2}e_{\bar{1},2,3}
+\frac{\gamma}{2}e_{1\bar{2},3} \label{eq:twoanti1} \\
&&e_{2,\bar{1},\bar{2},3}=e_{2,3}-e_{1,2}+\frac{\gamma}{2}e_{\bar{1},2,3}
-\frac{\gamma}{2}e_{1,\bar{2},3} \label{eq:twoanti2} \\
&&e_{\bar{2},\bar{1},2,3}=e_{2,3}+e_{1,2}+\frac{\gamma}{2}e_{\bar{1},2,3}
-\frac{\gamma}{2}e_{1,\bar{2},3} \label{eq:twoanti3} \\
&&e_{\bar{2},1,\bar{2},3}=e_{2,3}+e_{1,2}-\frac{\gamma}{2}e_{\bar{1},2,3}
+\frac{\gamma}{2}e_{1,\bar{2},3}\,,\label{eq:twoanti4}
\eea
and lastly, the two simple relations
\bea
&&e_{1,1,2}=\gamma e_{\bar{1},2}-2e_1 \\
&&e_{1,\bar{1},2}=\gamma e_{1,2}\,.
\eea
The above relations are sufficient for our calculations, because all terms
$e_{i_1i_2\ldots i_k\ldots i_l}$ in $T^N_{k,l}$ have index values ordered as
$ i_k<i_{k+1}$. When commuting with an element $e_k$ there will be at most
four index values involved, as can be seen from the Temperley-Lieb algebra
(\ref{Temperley}), because $e_k$ will commute with all generators $e_i$ appearing in
$e_{1,\ldots ,{k-1},k,{k+1},\ldots, N}$, except $e_{k-1}$ and $e_{k+1}$. When we
commute $e_k$ with a generic term
$e_{1,\ldots ,\widetilde{k-1},\tilde{k},\widetilde{k+1},\ldots, N}$
($\tilde{k}$ denotes $k$ or $\bar{k}$), we get
\[
[e_k,e_{1,\ldots ,\widetilde{k-1},\tilde{k},\widetilde{k+1},\ldots, N}]=
e_{1,\ldots ,k,\widetilde{k-1},\tilde{k},\widetilde{k+1},\ldots, N}\,.
\]

In each term of the quantity $T^N_{k,l}$, each generator $e_i$ appears at most
once. Looking at equations (\ref{eq:oneanti1}-\ref{eq:oneanti3}) and
(\ref{eq:twoanti1}-\ref{eq:twoanti4}), the first set of equations give rise to
terms for which this is not the case, because these contain the term $e_{B_2}$.
It is not possible to rewrite it so that each generator appears at most
once.

Thus, when commuting $Q_N$ with the boost operator, the only combinations
of barred indices which can give rise to an undesirable term are
\bea
&&[ne_n,e_{1,\ldots,\overline{n-1},n,n+1,\ldots} ]=
ne_{1,\ldots,n,\overline{n-1},n,n+1,\ldots} \label{eq:boost_oneanti1} \\
&&[ne_n,e_{1,\ldots,n-1,\bar{n},n+1,\ldots} ]=
ne_{1,\ldots,n,n-1,\bar{n},n+1,\ldots}.\label{eq:boost_oneanti2}
\eea
These two terms in $Q_N$ belong to the same $T^N_{k,l}$, which means they
have the same prefactor. From the relative minus signs in equations
(\ref{eq:oneanti1}-\ref{eq:oneanti2}), we see that the $e_{B_2}$ terms
cancel when adding the two quantities on the right hand sides of
(\ref{eq:boost_oneanti1}-\ref{eq:boost_oneanti2}).
In order for the sum above not to vanish we need to have an anti-commutator
either connecting to the right (a bar over $n+1$),
or the left (a bar over the index left of $n$, say it is $n-2$).
Assume that it is to the left. Then the sum of the two terms above is
\be
\label{eq:boost_twosum}
2ne_{1,\ldots,\overline{n-2},\bar{n},n+1,\ldots}\,.
\ee
We can also obtain this type of term from the commutators
\be
[(n-2)e_{n-2},e_{1,\ldots,n-3,n-2,\overline{n-1},\overline{n},\ldots}] = 
(n-2)e_{1,\ldots,n-2,n-3,n-2,\overline{n-1},\overline{n},\ldots}
\label{eq:boost_twoanti1}
\ee
\be
-\frac{2}{\gamma}
[ne_n,e_{1,\ldots,\overline{n-2},n,n+1,\ldots}] 
=-\frac{2}{\gamma} ne_{1,\ldots,\overline{n-2},n,n,n+1,\ldots}\,,\label{
eq:boost_twoanti2}
\ee
\be
-\frac{2}{\gamma}
[(n+1)e_{n+1},e_{1,\ldots,\overline{n-2},n,n+1,\ldots}] 
=-\frac{2}{\gamma} (n+1)e_{1,\ldots,\overline{n-2},n+1,n,n+1,\ldots}\,.\label{
eq:boost_twoanti3}
\ee
The factor $2/\gamma$ in the product with one gap and one bar 
is important in order for the $n$ to cancel, and it will be of
importance for the appearance of the general charge (see
equation (\ref{eq:Tdef})).
Once again, using the relations (\ref{eq:oneanti1}-\ref{eq:oneanti3}), we
add the three terms above and get (among other things)
\be
-(2n-3)e_{1,\ldots,\overline{n-2},\bar{n},n+1,\ldots}\,.
\label{eq:boost_twoanti_sum}
\ee
Thus we will get both the contributions
(\ref{eq:boost_twosum}) and
(\ref{eq:boost_twoanti_sum}).
Adding them up, it is clear that we get these terms symmetrically 
distributed with
no $n$-dependence. In this way we see that we obtain the correct structure.
We also know from the existence of the boost operator that the term
with the highest range number is $e_{1,2,3,\ldots,N}$.

Now we do the last preparation for what is needed to obtain
the exact form of the commuting charges. First, we introduce the notation
\be
\label{eq:Tdef}
\tilde{T}^N_k=\sum_{j=0}^{k-1}
\left(-\frac{\gamma^{-j}}{2}\right) T^N_{k-j,j}\,.
\ee

After some lengthy calculation, we can show that the following
relation holds, valid when $1<k<N-4$:
\be
0\stackrel{N-1}{=}
\big[\tilde{T}^1_0,\tilde{T}^N_k+
\frac{\gamma}{2}(\tilde{T}^{N-1}_{k+1}
-\tilde{T}^{N-1}_{k-1}) 
+\frac{\gamma^2}{2^2}(\tilde{T}^{N-2}_{k+2}
-2\tilde{T}^{N-2}_k
+\tilde{T}^{N-2}_{k-2})\big]\,.
\ee
The $N-1$ above the equality sign means that only terms with range number
$N-1$ are equal.
 When $k=1$,
\be
0\stackrel{N-1}{=}
\big[\tilde{T}^1_0,\tilde{T}^N_1+
\frac{\gamma}{2}\tilde{T}^{N-1}_{2} 
+\frac{\gamma^2}{2^2}(\tilde{T}^{N-2}_{3}
-\tilde{T}^{N-2}_1)\big]\,.
\ee
For $k=N-4$ and $k=N-3$, we have
\be
0\stackrel{N-1}{=}
\big[\tilde{T}^1_0,\tilde{T}^N_k+
\frac{\gamma}{2}(\tilde{T}^{N-1}_{k+1}
-\tilde{T}^{N-1}_{k-1}) 
+\frac{\gamma^2}{2^2}(
-2\tilde{T}^{N-2}_k
+\tilde{T}^{N-2}_{k-2})\big]\,,
\ee
and finally, for $k=N-2$ and $k=N-1$,
\be
0\stackrel{N-1}{=}
\big[\tilde{T}^1_0,\tilde{T}^N_k+
\frac{\gamma}{2}(\tilde{T}^{N-1}_{k+1}
-\tilde{T}^{N-1}_{k-1}) 
+\frac{\gamma^2}{2^2}
\tilde{T}^{N-2}_{k-2}\big]\,.
\ee
 We also have the relations
\be
0\stackrel{N-1}{=}[\tilde{T}^1_0,\tilde{T}^N_1+
\frac{\gamma}{2}\tilde{T}^{N-1}_2],
\ee
and
\be
0=[\tilde{T}^1_0,\tilde{T}^N_0]+
\frac{\gamma}{2}[\tilde{T}^1_0,\tilde{T}^{N-1}_1]\big|_{N},
\ee
where $|_N$ on the right side illustrates that the left side
is only equal to the term with range $N$ on the right side.
From this, we can deduce the following expression for the commuting
charges\footnote{$\lfloor m\rfloor$ is the integer part of $m$,
and $a\vee b := \max(a,b)$}:
\be
Q_N=\sum_{n=0}^{N-2}\left(\sum_{l=0\vee W_{N,n}}^{\lfloor n/2\rfloor}
 \frac{\gamma^n}{2^n}C_{n,l}\tilde{T}^{N-n}_{n-2l}\right),
\qquad W_{N,n}=
\left\{
\begin{array}{cc}
-\frac{N-2n-3}{2} &,N\;\mathrm{odd} \\
-\frac{N-2n-2}{2} &,N\;\mathrm{even},
\end{array}
\right.
\ee
where the coefficients are given by (see Appendix)
\be
C_{n,l}=(-1)^l\left(
\left(\begin{array}{c}
n \\ l
\end{array}\right)-\sum_{j=1}^{l}
\left(\begin{array}{c}
n-j \\ l-j
\end{array}\right) \right),
\ee
for $n>1$ and $C_{0,0}=1$ and $C_{1,0}=1$.

\section{Conclusion}
We have obtained an expression for a generic conserved charge of a
local Hamiltonian satisfying the Temperley-Lieb algebra,
in a simple form. This expression can be used to verify
whether the generalisation, which we have given here to the first
few orders in $\lambda$, to the higher loop $N=4$ model, is integrable.
We expect that when adding higher loop contributions, the charges can
also be expressed in a similar simple form.
There are two separate questions we would like to have an answer to.

Firstly, is the Hamiltonian (\ref{Hamiltonian2}) integrable when including
all orders in the parameter $\lambda$?
And secondly, this Hamiltonian has the correct interaction range to
given orders in $\lambda$ to describe a field theory, but will the
dilatation operator of the $q$-deformed theory have the correct
 coefficients in front of the different interaction terms to be described
by our model?
We have seen that simply demanding the existence of one commuting
charge puts strong conditions on the appearance of the field theory.
It is enough for us to do a three loop calculation on the field theory side
to be able to rule out integrability if the coefficients do not come
out as in equation (\ref{Hamiltonian2}) modulo lower charges and shift in
$\lambda$. If the three-loop calculation would come out in favour of our
model then it would be very interesting if someone could do a four-loop
calculation.

We would like to emphasise that the model is valid for any 
representation of the Temperley-Lieb algebra, and as such could very well 
describe the matrix of anomalous dimensions of some other field theory.
Another interesting question is if these types of long-range spin chains
always can be constructed as effective theories of Hubbard type, as is the
case for the dilatation operator of $\mathcal{N}=4$ SYM.

To summarise, we have not yet revealed the infinite tower of commuting
charges in the higher loop deformed case, but we believe
that not many days remain before
they will be displayed.
\subsubsection*{Acknowledgments}
We would like to thank Sergey Frolov and Matthias Staudacher 
 for interesting discussions.
This work was supported by the Alexander von Humboldt foundation.
\appendix
\section{Some higher charges}
Here we have collected the five first of the two loop charges:
\bea
Q_1&=\tilde{T}^1_{0}+\lambda  \tilde{T}^2_{1} \nonumber \\
Q_2&=\tilde{T}^2_{0}+ 
\lambda \, \tilde{T}^3_{1}
 \nonumber \\
Q_3&=\tilde{T}^3_{0}+\frac{\gamma}{2}\tilde{T}^2_{1}+ 
\lambda \left(T^4_{1,0}+\gamma T^3_{2,0}-T^3_{1,1}
-2TB^3_{0,0}-(\frac{\gamma^2}{2}-4)T^2_{1,0} \right)
 \nonumber \\
Q_4&=\tilde{T}^4_0+\frac{\gamma}{2}\tilde{T}^3_1
+\lambda \left(T^5_{1,0}+\gamma T^4_{2,0}-T^4_{1,1}-2TB^4_{0,0}-
(\frac{\gamma^2}{2}-3)T^3_{1,0}\right)
\nonumber \\
Q_5&=\tilde{T}^5_{0}+\frac{\gamma}{2}\tilde{T}^4_{1}+
\frac{\gamma^2}{2}(\tilde{T}^3_{2}+\tilde{T}^3_{0})+ 
\lambda \left( T^6_{1,0}+\gamma \,T^5_{2,0}-T^5_{1,1}
+\frac{3\gamma^2}{4}T^4_{3,0}-\gamma T^4_{2,1}+T^4_{1,2}
\right.
 \nonumber \\
& -2 TB^5_{0,0}-\gamma TB^4_{1,0}-2 TB^3_{0,0}
+e_{\bar{i},i+1,i+2,i+3}+e_{i,i+1,\overline{i+2},i+3}
 \nonumber \\
&\left. +\frac{2}{\gamma}\left(-2+\frac{\gamma^2}{4}\right)T^5_{0,0}+
\left(\frac{3\gamma}{2}-\frac{3\gamma^3}{8}\right)T^3_{2,0}+
\left(3-\frac{3\gamma^2}{4}+\frac{3\gamma^4}{16}\right)T^2_{1,0}
+\frac{3\gamma^2}{4}T^3_{1,1}\right)\,.
 \nonumber
\eea
The charges are only specified modulo lower charges.

\section{The coefficients in the generic charge}
\begin{figure}
\parbox{7cm}{\centering\includegraphics[height=5.5cm]{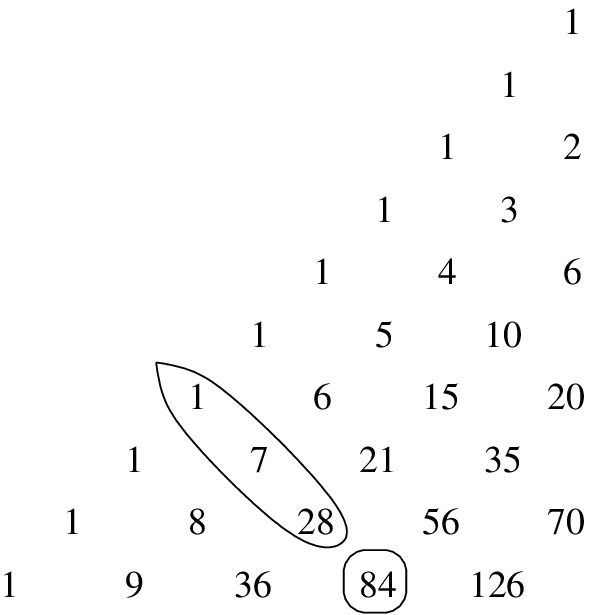}}
\parbox{7cm}{\centering\includegraphics[height=5.5cm]{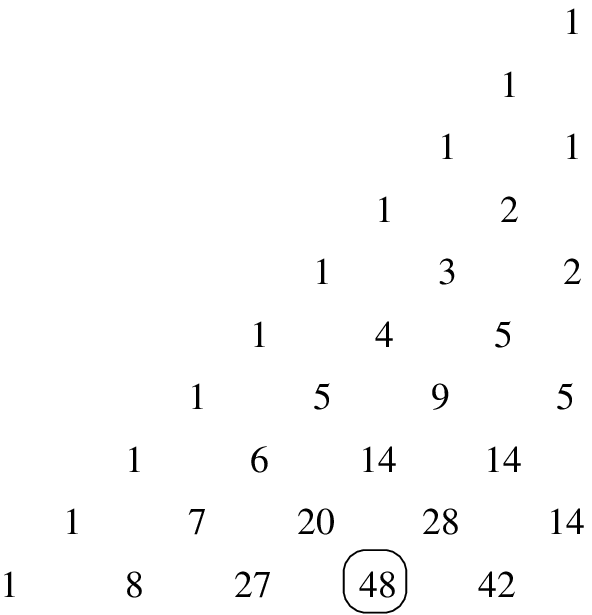}}
\caption{Pascal's triangle together with the coefficient triangle.}
\label{fig:pascals}
\end{figure}

In figure \ref{fig:pascals}, we have pictured Pascal's triangle
together with the triangle of coefficients that we get for our charges. Our
coefficient triangle can be obtained by adding up several of Pascal's
triangles in such a way that the new triangle will have only zeroes on
the right hand side (in the picture we have not displayed the right hand
side of Pascal's triangle). As an example, we have marked the coefficient
value $48$ in our triangle. It can be obtained from Pascal's triangle by
taking $84$ minus the sum of the marked diagonal above $84$.

We can simply read off the values from the triangle when writing down one
of the charges, e.g. charge eight:
$$
\fl Q_8=\tilde{T}^8_0+\frac{\gamma}{2}T^7_1 
+\frac{\gamma^2}{2^2}(T^6_2-T^6_0) 
+\frac{\gamma^3}{2^3}(T^5_3-2T^5_1) 
+\frac{\gamma^4}{2^4}(-3T^4_2+2T^4_0)
+\frac{\gamma^5}{2^5}(5T^3_1)\,.
$$

\section*{References}
\bibliographystyle{hunsrt}
\bibliography{leighref2}

\end{document}